\documentclass{appolb}
\usepackage{graphicx,amssymb,amsmath}

\usepackage{lineno}

\begin{document}
\title{Charm and beauty production and hadronization with the ALICE experiment
\thanks{Presented at BEACH 2022}%
}
\author{R\'obert V\'ertesi (for the ALICE Collaboration)
\address{Wigner RCP, Budapest, Hungary}
}
\maketitle
\begin{abstract}
This contribution summarizes a selection of recent results from the ALICE experiment. We show the cross-section of beauty jets in pp collisions, as well as the charmed baryon-to-meson production ratios, down to unprecedentedly low momenta, at $\sqrt{s_{\rm NN}} = 5.02$ and 13 TeV collision energies. We present the nuclear modification of ${\rm D}^0$-jets and the $v_2$ of non-prompt ${\rm D}^0$ in Pb-Pb collisions at $\sqrt{s_{\rm NN}} = 5.02$ TeV. We also discuss the implications and constraints from comparing these data to some of the most recent theoretical models.

\end{abstract}

\section{Introduction}

Heavy-flavor quarks (charm and beauty) are produced in early partonic processes of high-energy collisions, thus they provide a unique opportunity to probe the entire evolution of the strongly interacting system~\cite{Andronic:2015wma}. Heavy-flavor production in proton--proton (pp) collisions serves as fundamental benchmark for quantum-chromodynamics (QCD) calculations and fragmentation in the mesonic and baryonic sector. 
The production of heavy flavor in small collision systems is generally described using the factorization approach, 
where the production cross-section of a certain hadron is described in terms of three independent factors: {\it i)} the parton distribution functions of the incoming hadrons, {\it ii)} the cross section of the elementary parton-parton scattering in which the heavy quarks are produced, and {\it iii)} the heavy-quark fragmentation function into the given hadron.
Since the mass of heavy quarks is significantly above the QCD energy scale ($m_{b,c} \gg \Lambda_{\rm QCD})$, 
their production can be used to test perturbative QCD models down to low momenta. 
In contrary to the light jets which are mostly initiated by gluons, heavy-flavor jets are initiated by quarks (c or b). By the comparison of their production the effect of different color-charges of quarks and gluons can be tested. On the other hand, fragmentation of massive heavy quarks is generally harder because of the dead-cone effect~\cite{Dokshitzer:1991fd,ALICE:2021aqk}. 
%
%
In addition, heavy-flavor probes in proton on heavy-ion (p--A) collisions test cold nuclear matter effects. 
Conversely, heavy-ion (A--A) collisions allow for the detailed mapping of transport properties, collective motion and the thermalization of the quark-gluon plasma (QGP). 

In the measurements recapitulated here, heavy flavor hadrons are reconstructed via their hadronic decay channels using the ALICE Inner Tracking System (ITS) and the Time Projection Chamber (TPC) at central rapidity ($|\eta|<0.9$). Electrons from the semi-leptonic decays are identified with the Electromagnetic Calorimeter. Since the lifetime of heavy-flavor is long, the sample purity is enhanced using the ITS based on the identification of the displaced decay vertex ($c\tau_c \approx 100 - 300\ \mu$m, $c\tau_b \approx 400 - 500\ \mu$m compared to the ITS secondary vertex resolution that is less than 100\ $\mu$m)~\cite{ALICE:1999cls}. Charged-particle identification is done in the TPC using specific energy loss (${\rm d}E/{\rm d}x$) aided by precise time-of-flight measurements by the TOF detector, while the V0 detectors at forward rapidity are used for event characterization. A detailed description of the ALICE apparatus can be found elsewhere~\cite{ALICE:2014sbx}.

\section{Heavy-flavor production in proton-proton collisions}

\begin{figure}[htb]
\vspace{-5mm}
\centerline{%
\begin{minipage}{6in}
  \centering
  \raisebox{-0.5\height}{\includegraphics[width=.37\columnwidth]{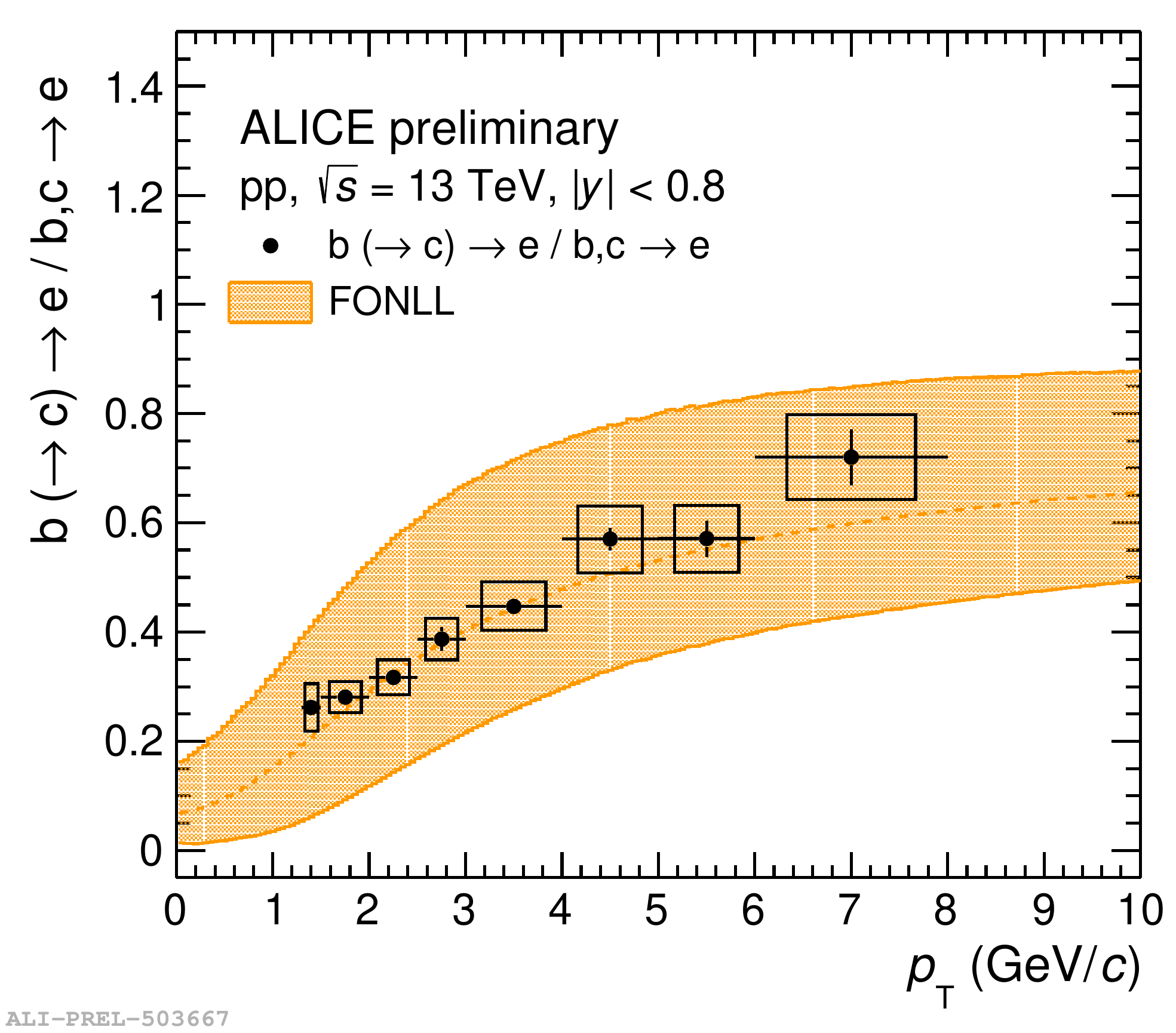}}
  \hspace*{.02\columnwidth}
  \raisebox{-0.5\height}{\includegraphics[width=.42\columnwidth]{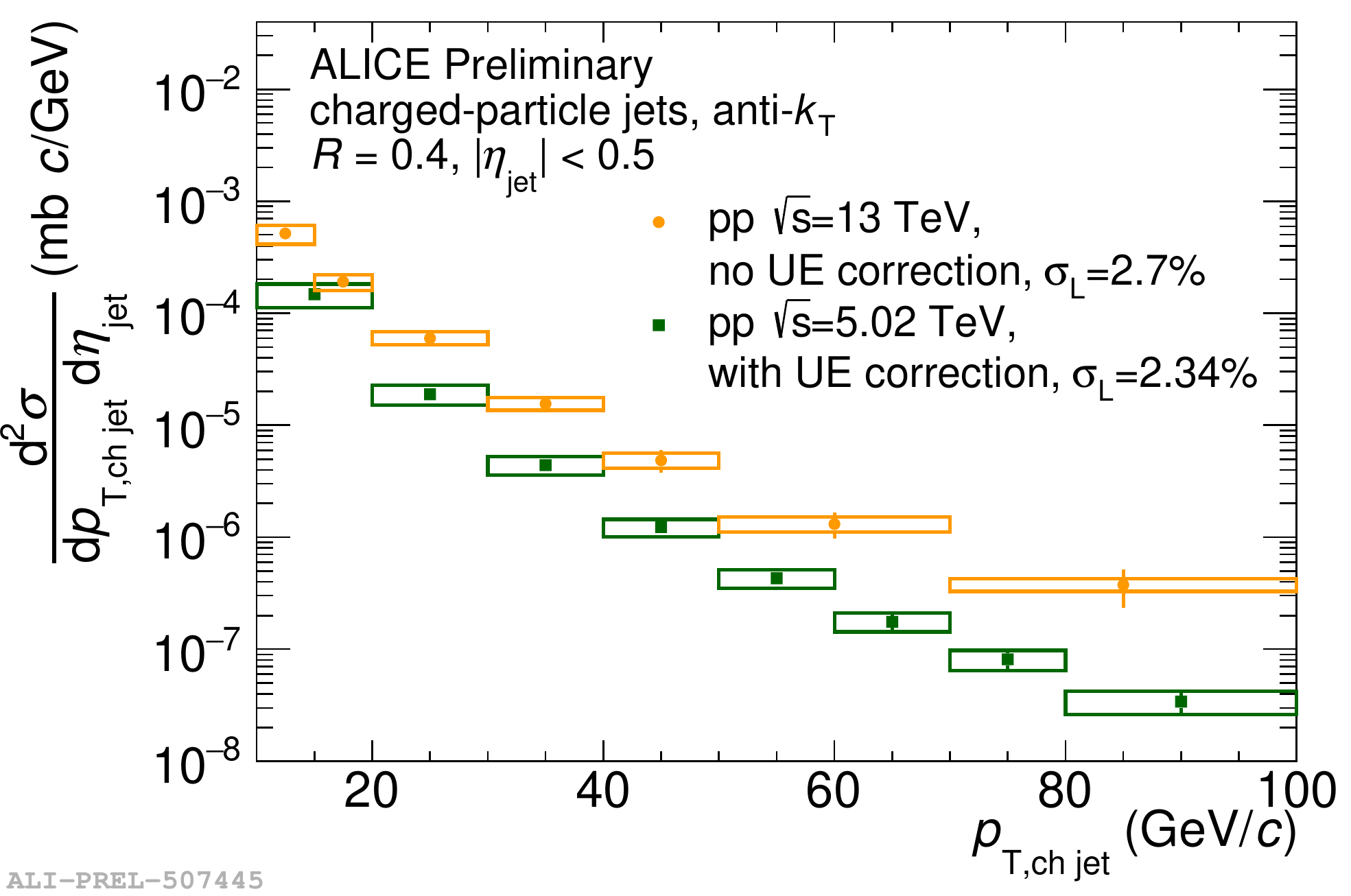}}
\end{minipage}
}
\caption{(Left) The ratio of electrons from beauty-hadron decays to those from heavy-flavor hadron decays in pp collisions at $\sqrt{s}=13$~TeV, for $1.3 < p_{\rm T} < 8$ GeV/$c$. (Right) Comparison of the fully corrected b-jet spectra for pp collisions at $\sqrt{s}=13$~TeV and 5.02 TeV.}
\label{Fig:hfprod}
\end{figure}

ALICE has measured the production of both the prompt and the non-prompt (beauty-hadron decay) contributions to the non-strange ${\rm D}^0$ and ${\rm D}^+$ meson yields down to very low momenta ($p_{\rm T} \approx 0$), as well as the yields of the strange ${\rm D}_s^+$~\cite{ALICE:2015ikl}.
The yields of both charm and beauty hadrons are typically well-described by several models based on factorization~\cite{Cacciari:2001td, Helenius:2018uul}, although GM-VFNS tends to underestimate non-prompt D-meson yields.
The left panel in Fig.~\ref{Fig:hfprod} shows the fraction of heavy-flavor decay electrons stemming from beauty hadrons, compared to FONLL~\cite{Cacciari:2001td} calculations.
Currently available data on heavy-flavor meson and electron production already provide strong constraints for models.

Jets containing heavy flavor serve as a direct proxy for the heavy quark. In ALICE, charm jets are identified by tagging a ${\rm D}^0$ meson within a jet~\cite{ALICE:2022mur}, while b-jets are tagged using different methods based on the displacement of secondary vertex~\cite{ALICE:2021wct}. The current data already strongly restrict models and provide a unique opportunity to study flavor-dependent jet properties. The right panel of Fig.~\ref{Fig:hfprod} shows the cross section of b-jet production at $\sqrt{s}=5.02$ TeV and 13 TeV. A hardening of the $p_{\rm T}$-spectrum can be observed at higher collision energies.

\begin{figure}[htb]
\vspace{-5mm}
\centerline{%
\begin{minipage}{6in}
  \centering
  \raisebox{-0.5\height}{\includegraphics[width=.37\columnwidth]{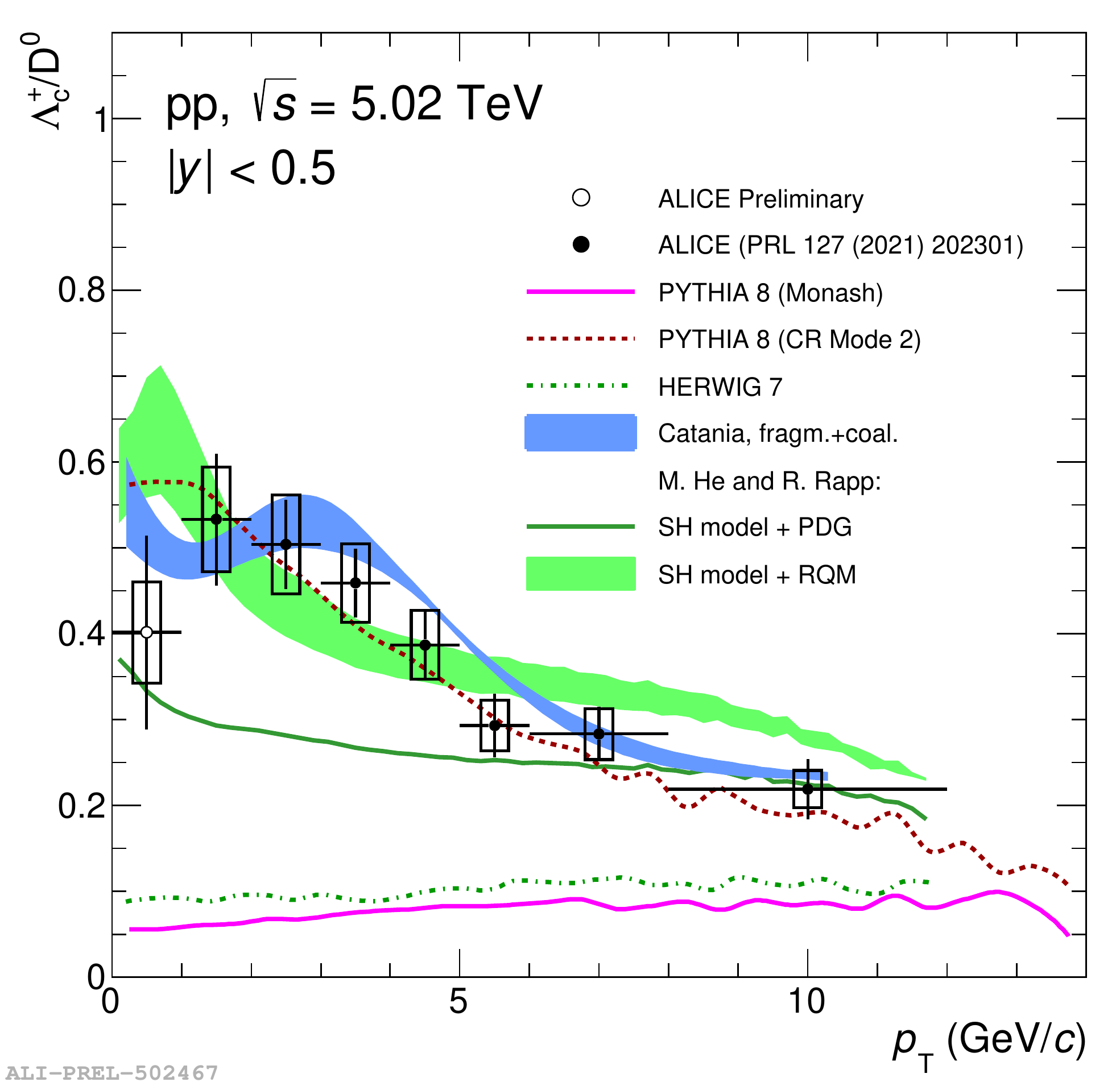}}
  \hspace*{.05\columnwidth}
  \raisebox{-0.5\height}{\includegraphics[width=.42\columnwidth]{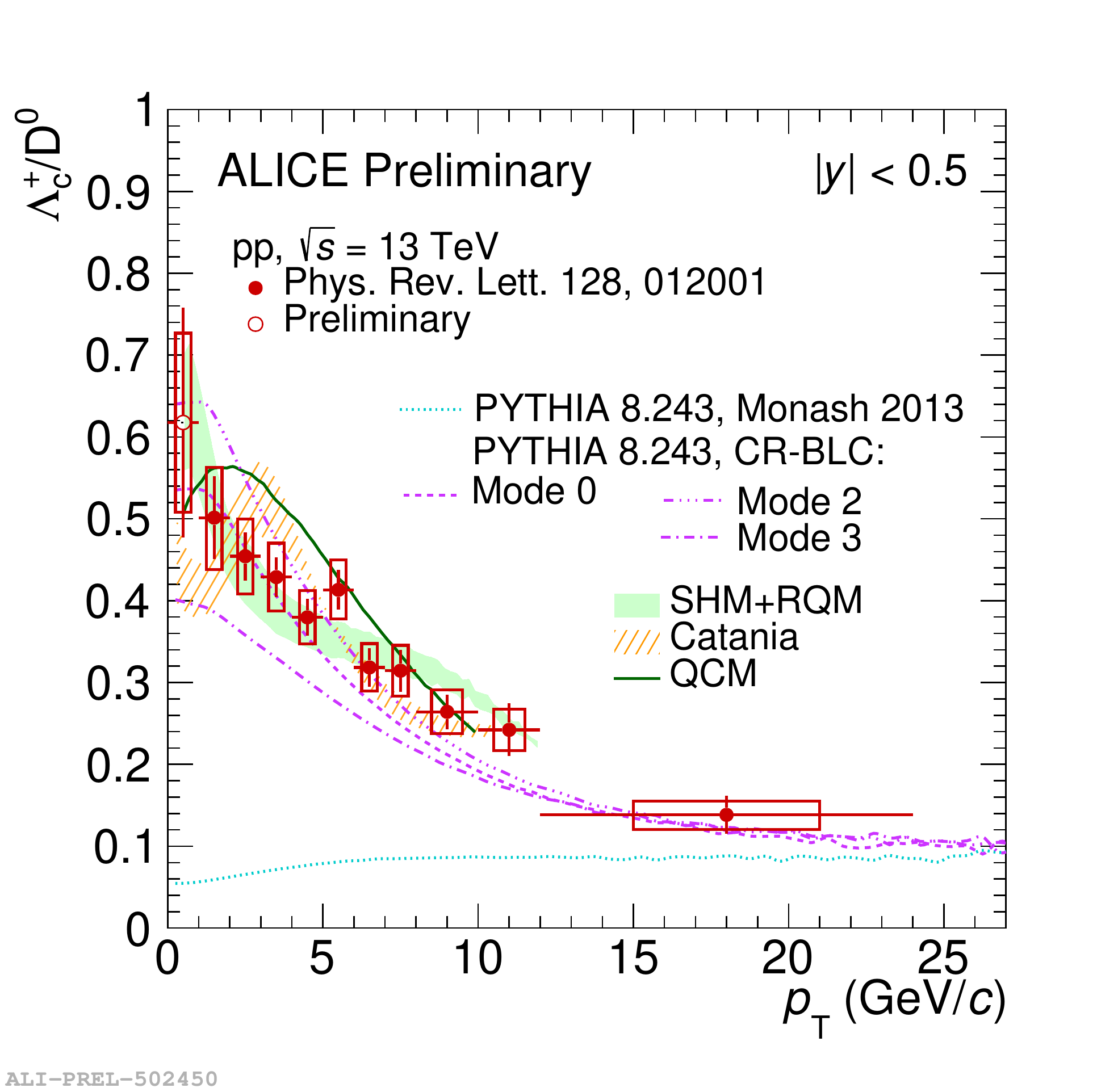}}
\end{minipage}
}
\caption{The $\Lambda_c^+/{\rm D}^0$ baryon-to-meson ratio in pp collisions at $\sqrt{s}=5.02$ TeV (left) and 13 TeV (right), down to $p_{\rm T} \approx 0$, compared with model calculations~\cite{Christiansen:2015yqa,Plumari:2017ntm,Song:2018tpv,He:2019tik}.}
\label{Fig:lambdac}
\end{figure}

The comparison of charm-baryon production to that of charm mesons is specific to charm hadronization. ALICE measured the $\Lambda_c^+ / D^0$ ratios down to $p_{\rm T} \approx 0$, as shown on Fig.~\ref{Fig:lambdac} for collisions at $\sqrt{s}=5.02$ TeV as well as 13 TeV.

The ratios are underestimated by models based on factorization approach with fragmentation functions from $e^+ e^-$ collisions~\cite{Sjostrand:2014zea}, which suggests that the universality of heavy-flavor fragmentation is broken in hadronic collisions. 
While several scenarios have been suggested, including color reconnection beyond leading order~\cite{Christiansen:2015yqa}, coalescence of charm and light quarks~\cite{Plumari:2017ntm,Song:2018tpv}, and the presence of an augmented set of charmed resonances~\cite{He:2019tik}, the current understanding is far from complete, as heavier charm states ($\Xi_c^{0,+}$, $\Sigma_c^{0,++}$, $\Omega_c^{0}$) still pose a challenge for these models~\cite{ALICE:2021rzj,ALICE:2019pid,ALICE:2022cop}. Future event-shape-dependent measurements as well as heavy-flavor jet and correlation measurements will help clarify this question~\cite{Varga:2021jzb}.

\section{Toward larger systems}

\begin{figure}[htb]
\vspace{-5mm}
\centerline{%
\begin{minipage}{6in}
  \centering
  \raisebox{-0.5\height}{\includegraphics[width=.38\columnwidth]{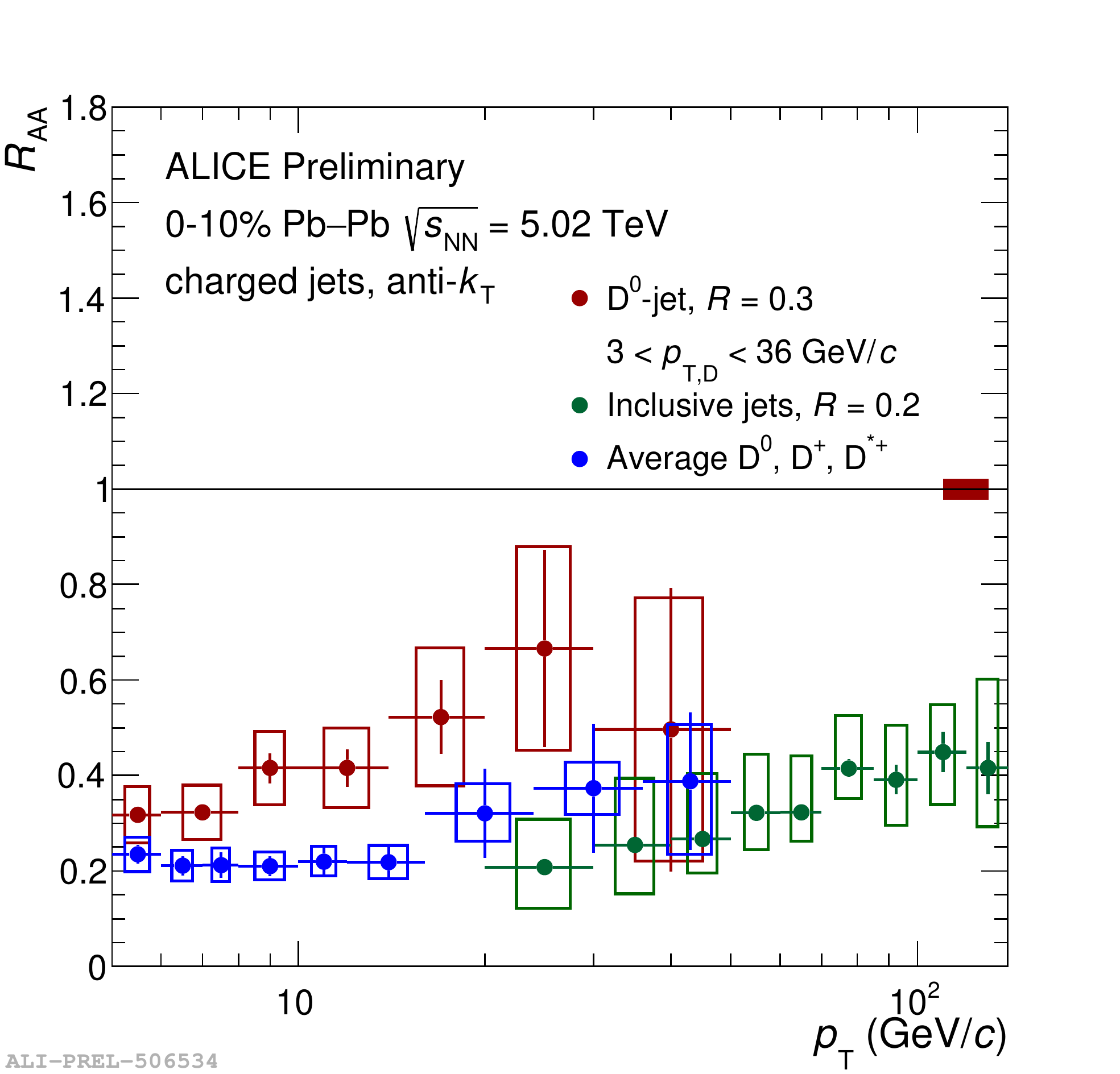}}
  \hspace*{.04\columnwidth}
  \raisebox{-0.55\height}{\includegraphics[width=.37\columnwidth]{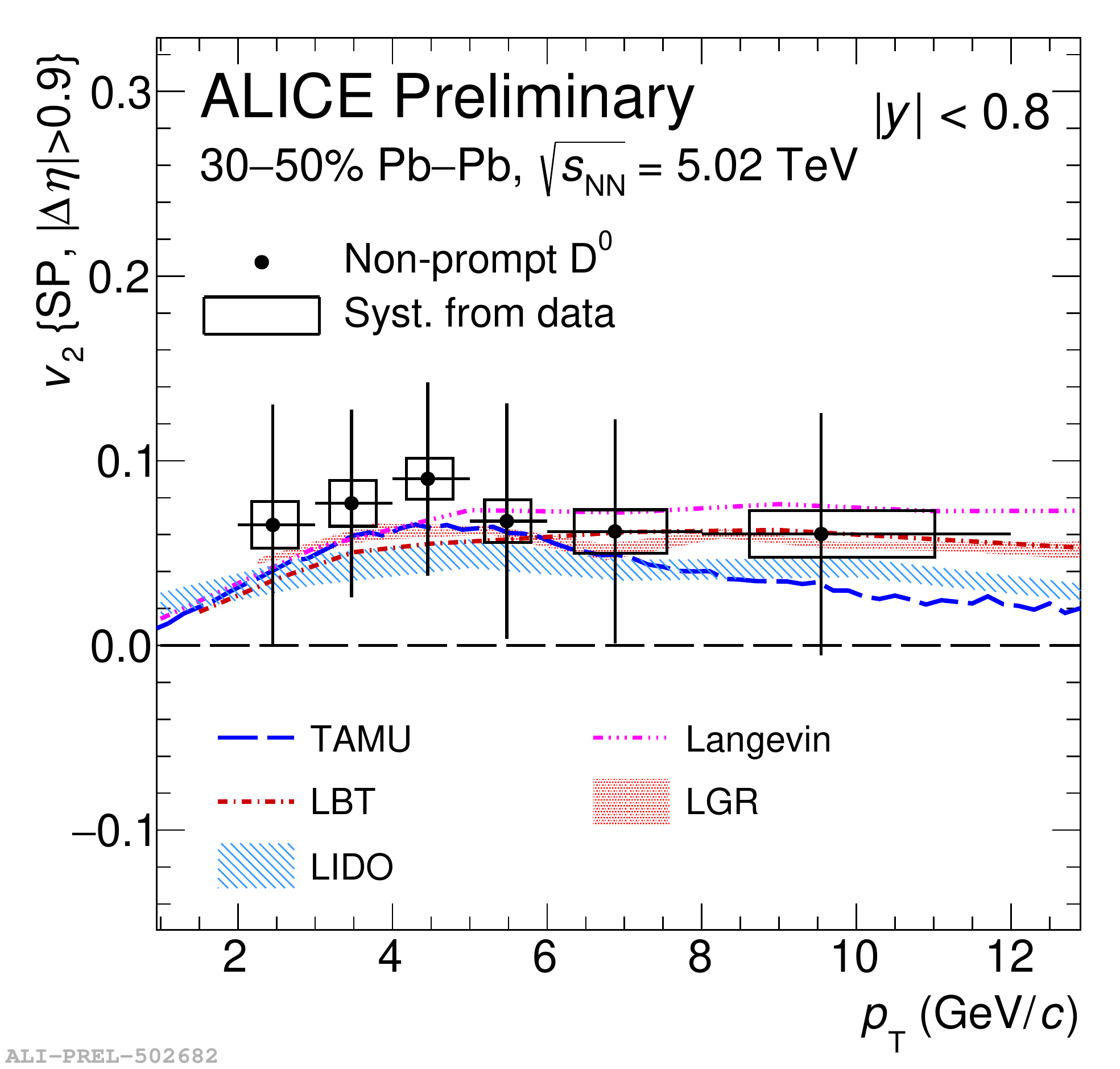}}
\end{minipage}
}
\caption{(Left) The $R_{\rm AA}$ of ${\rm D}^0$ jets compared to that of inclusive charged-particle jets and the average of D-mesons in Pb--Pb events from the in 0--10\% centrality class at $\sqrt{s}=5.02$ TeV. (Right) Non-prompt ${\rm D}^0$ $v_2$ in Pb--Pb collisions at $\sqrt{s}=5.02$ TeV in the 30--50\% centrality class.}
\label{Fig:AAcoll}
\end{figure}

Heavy-flavor measurements in larger (p--A and A--A) systems probes the energy loss and collectivity effects caused by hot and cold nuclear matter.
In p--Pb collisions, cold nuclear effects are not significant by the current experimental precision~\cite{ALICE:2021wct}. On the other hand, a significant nuclear modification in case of Pb--Pb collisions. Fig.~\ref{Fig:AAcoll} (left) shows the nuclear modification factor $R_{\rm AA}$ for $D^0$-tagged jets at $\sqrt{s}=5.02$ TeV.
The ${\rm D}^0$ jets are less suppressed than ${\rm D}^0$ mesons, which may be caused by the sensitivity of jets to different energy
loss of quarks and gluons (color-charge effect), and possibly to the mass (dead cone) effect. Also note that the measurements of both the  ${\rm D}_s^+$ meson and the $\Lambda_c^+$ baryon indicate a higher $R_{\rm AA}$ value at intermediate $p_{\rm T}$ values, which suggests that the 
hadronization mechanisms are also modified by the QGP~\cite{ALICE:2021bib}. 

In the $p_{\rm T} \lesssim 8$ GeV/$c$ range the non-strange D-meson $R_{\rm AA}$ is significantly higher than that of light hadrons. On the other hand, D mesons exhibit positive elliptic flow $v_2$~\cite{ALICE:2021rxa}. These two observations can be simultaneously understood by transport models that assume charm--light quark coalescence and require that charm quarks take part in the collective motion~\cite{Song:2015sfa,Beraudo:2014boa}. Another model interprets the results with a modified Boltzmann and radiative energy loss~\cite{Ke:2018jem}. Fig.~\ref{Fig:AAcoll} (right) shows the $v_2$ of non-prompt ${\rm D}^0$ mesons from beauty decays. These results are consistent with model calculations (e.g. in \cite{Ke:2018jem}) that predict a $v_2$ for beauty that is non-zero but substantially smaller than charm $v_2$. More precise data will provide important constraints on the extent to which beauty participates in the collective motion.

\section{Summary and outlook}

ALICE performed a set of measurements of the production and fragmentation of charm and beauty in pp collisions. These include the first direct measurement of the QCD dead cone, and serve as precision pQCD benchmark as well as tests of parton shower and fragmentation via jet substructure measurements. Detailed studies of fragmentation with charmed mesons and baryons showed that fragmentation is not universal. 
In p--Pb collisions, cold nuclear effects do not play a strong role in case of heavy flavor. 
Measurement of the nuclear modification and collectivity of different charm hadrons in heavy-ion collisions can be interpreted with models based on charm-light quark coalescence. The first low-$p_{\rm T}$ D-jet measurement in Pb-Pb shows less suppression than for D mesons. From the recently started Run 3 phase we expect two orders of magnitude more minimum-bias data with upgraded detectors, which will allow for unprecedented differential measurements and provide us with a golden opportunity to study the charm-baryon sector and beauty production.

This work has been supported by the NKFIH grants OTKA FK131979 and K135515, as well as by the 2019-2.1.11-T\'ET-2019-00078 and 2019-2.1.6-NEMZ\_KI-2019-00011 projects.


\vspace{-3mm}
\bibliographystyle{ieeetr}
\bibliography{myrefs.bib}

\end{document}